\documentclass{aa}  
\usepackage{graphicx}
\usepackage{txfonts}
\usepackage{color}
\usepackage{siunitx}
\usepackage{upgreek}
\usepackage[colorlinks,citecolor=blue,urlcolor=blue,filecolor=blue,linkcolor=blue]{hyperref}

\begin{document} 
    \title{Irradiated but not eclipsed, the case of PSR\,J0610$-$2100}

\author{
  E.\ van der Wateren\inst{\ref{astron}, \ref{nijmegen}} \fnmsep\thanks{\url{emma.vanderwateren@gmail.com}}
  \and
  C.\ G.\ Bassa\inst{\ref{astron}}
  \and
  C.\ J.\ Clark\inst{\ref{jb}, \ref{aei}, \ref{leibniz}}
  \and
  R.\ P.\ Breton\inst{\ref{jb}}
  \and
  I.\ Cognard\inst{\ref{lpc2e}, \ref{nancay}}
  \and
  L.\ Guillemot\inst{\ref{lpc2e}, \ref{nancay}}
  \and
  G.\ H.\ Janssen\inst{\ref{astron}, \ref{nijmegen}}
  \and \\
  A.\ G.\ Lyne\inst{\ref{jb}}
  \and
  B.\ W.\ Stappers\inst{\ref{jb}}
  \and
  G.\ Theureau\inst{\ref{lpc2e}, \ref{nancay}, \ref{luth}}
}

\institute{
  ASTRON, Netherlands Institute for Radio Astronomy, Oude
  Hoogeveensedijk 4, 7991 PD Dwingeloo, The Netherlands\label{astron}
  \and
  Department of Astrophysics/IMAPP, Radboud University Nijmegen,
  P.O. Box 9010, 6500 GL Nijmegen, The Netherlands\label{nijmegen}
  \and  
  Jodrell Bank Centre for Astrophysics, Department of Physics and
  Astronomy, University of Manchester, Manchester M13 9PL,
  UK\label{jb}
  \and
  Max-Planck-Institut f\"{u}r
  Gravitationsphysik (Albert-Einstein-Institut), 30167 Hannover, Germany\label{aei}
  \and
  Leibniz Universit\"{a}t Hannover, 30167 Hannover, Germany\label{leibniz}
  \and
  Laboratoire de Physique et Chimie de l'Environnement et de l'Espace,
  Universit\'{e} d'Orl\'{e}ans / CNRS, 45071 Orl\'{e}ans Cedex 02,
  France\label{lpc2e}
  \and
  Station de Radioastronomie de Nan\c{c}ay, Observatoire de Paris,
  CNRS/INSU, 18330 Nan\c{c}ay, France\label{nancay} 
  \and 
  Laboratoire Univers et Th\'eories, Observatoire de Paris, Universit\'e PSLU, CNRS, 92195 Meudon Principal Cedex, France\label{luth}
}

\date{Received November 25, 2021; accepted March 7, 2022}

\abstract{
  We report on radio timing observations of the black widow binary pulsar J0610$-$2100 and optical observations of its binary companion. The radio timing observations extend the timing baseline to 16\,yr and reveal a marginal detection of the orbital period derivative, but they show no significant evidence of orbital variations such as those seen in other black widow pulsars. Furthermore, no eclipses are seen in the observations at observing frequencies ranging from 310 to 2700\,MHz. The optical $V\!RI$ light curves were modulated with the orbital period, reaching maximum brightness of $V=26.8$, $R=25.4$, and $I=23.8$ at superior conjunction of the companion, confirming irradiation of the companion by the pulsar. Modelling the light curves indicates that the companion is likely not filling its Roche lobe, while having a moderate inclination ($i > 54\degr$). We find an unusually low temperature and a low irradiation for the irradiated hemisphere of the companion. We investigate the absence of radio eclipses in PSR\,J0610$-$2100 and in other black widow systems in relation to their binary, pulsar, and companion properties. We also discuss the suitability of PSR\,J0610$-$2100 for pulsar timing array observations aimed at detecting nano-Hertz gravitational waves.
}


\maketitle

\section{Introduction}
Amongst the radio pulsars residing in binaries, black widows are a class of binary millisecond pulsars with very low-mass binary companions ($M_\mathrm{c}\lesssim 0.05$\,M$_\odot$) in compact orbits (orbital period $P_\mathrm{b}\lesssim1$\,day) \citep{Roberts12,Chen13}. These binary systems display a set of observational phenomena not seen in the more typical binary millisecond pulsars with low-mass ($0.1$\,M$_\odot<M_\mathrm{c}<0.5$\,M$_\odot$) white dwarf companions in wider orbits ($P_\mathrm{b}>1$\,day). The most common phenomenon in black widows is the presence of radio eclipses, where the pulsed radio emission from the pulsar disappears over a fraction of the orbit as the binary companion moves close to the line of sight of the pulsar. However, radio eclipses are not seen in all black widows (e.g. \citealt{Breton2013, Stovall14}). For those black widow systems whose binary companions are detected at optical wavelengths, the companions display large sinusoidal variations in their brightness and colour, which is in phase with the binary orbit \citep{Stappers01a,Breton2013}. High precision pulsar timing of the millisecond pulsars in black widow systems reveal that some show variability of the orbital period on the timescales of months to years (e.g. \citealt{Arzoumanian94, Ng14, Shaifullah16}). Members of the class of redback binary millisecond pulsars display these phenomena as well, typically showing longer radio eclipses and larger orbital variations; these systems can be clearly distinguished from black widow systems due to their higher companion masses ($M_\mathrm{c}\simeq0.1 - 0.4$\,M$_\odot$; \citealt{Roberts12,Chen13}).

All of these observational phenomena are a result of the combination of the energetic wind from the radio pulsar ($\dot{E}\sim10^{34-35}$\,erg\,s$^{-1}$) and the compactness of the binary orbit. This wind irradiates the hemisphere of the binary companion facing the pulsar, leading to the observed optical and near-infrared variations. With the companions close to filling their Roche lobes, they can undergo wind-driven mass loss, leading to the observed radio eclipses. Finally, the orbital period variations are explained through magnetically, rotationally and/or tidally induced quadrupole distortions, leading to coupling between the binary orbit and the pulsar spin \citep{Applegate92, Applegate94,Voisin19}.

High precision pulsar timing of radio millisecond pulsars provides the possibility of detecting nano-Hertz gravitational waves with pulsar timing arrays (PTAs, \citealt{Foster90}). The sensitivity of PTAs strongly depends on the number of millisecond pulsars that can be included \citep{Siemens13}. Hence, black widow systems are observed by the European Pulsar Timing Array (EPTA; \citealt{Desvignes16}), the North American Nano-Hertz Observatory for Gravitational Waves (NANOgrav; \citealt{Arzoumanian18}), and the Parkes Pulsar Timing Array (PPTA; \citealt{Kerr20}) to determine whether their suitability is limited by orbital period variations (e.g. \citealt{Bak20}).

Many new radio millisecond pulsars in black widow systems have been discovered by targeting $\gamma$-ray sources discovered by the Fermi Gamma-Ray Space Telescope since 2009 \citep{Ray12}. Hence, long pulsar timing baselines are not yet available for these systems. The first two black widow systems discovered before Fermi, PSRs\,B1957+20 and J2051$-$0827, were discovered in 1988 \citep{Fruchter88} and 1995 \citep{Stappers96} and both exhibit orbital period variations \citep{Arzoumanian94, Doroshenko01}, radio eclipses \citep{Fruchter88,Stappers96} as well as optical variability of their binary companions \citep{Stappers99,Reynolds07}. The third system, PSR\,J0610$-$2100 discovered by \citet{Burgay06} in 2003, is a 3.86\,ms pulsar orbiting a $M_\mathrm{c}\approx0.02$\,M$_\odot$ companion in a 6.86-hour orbit. While the system shows optical variations \citep{Pallanca12}, high precision pulsar timing obtained to date does not report orbital variations or show evidence of radio eclipses \citep{Espinoza13,Desvignes16}. 

In this paper, we present radio observations of PSR\,J0610$-$2100 obtained with several radio telescopes. We combine the radio observations with optical observations of the companion of PSR\,J0610$-$2100 from the European Southern Observatory (ESO) to constrain the properties of this black widow system. The observations are described in \S\,\ref{observations} and the data analysis is presented in \S\,\ref{analysis}. We present our results in \S\,\ref{results} and discuss and conclude in \S\,\ref{discussion}.

\section{Observations}
\label{observations}
\subsection{Radio observations}
The radio observations presented in this paper were obtained with the 76-m Lovell telescope at the Jodrell Bank Observatory in the United Kingdom, the 94-m diameter equivalent Nan\c{c}ay Radio Telescope in France, the 64-m Parkes radio telescope in Australia, and the $14\times25$-m diameter Westerbork Synthesis Radio Telescope in the Netherlands. Together, these datasets cover a 16-year time span from the discovery of PSR\,J0610$-$2100 in 2003 May \citep{Burgay06} until 2019 June. The Parkes observations were obtained with an Analogue Filterbank (AFB) at a centre frequency of 1390\,MHz and the Parkes Digital Filterbanks (PDFB3 and PDFB4) at 1369\,MHz and consist of pulse times-of-arrival (TOAs) previously presented in \citet{Burgay06} and \citet{Espinoza13}. The Nan\c{c}ay observations consist of TOAs which have been published as part of the European Pulsar Timing Array (EPTA) Data Release 1.0 (DR1; \citealt{Desvignes16}). The observations were obtained with the Berkeley-Orleans-Nan\c{c}ay (BON) instrument at centre frequencies 1400, 1600, and 2000\,MHz, with bandwidths of 64 or 128\,MHz. 

These TOAs were supplemented with folded and dedispersed profiles (in the \textsc{psrchive} format; \citealt{Hotan04}) from Parkes, Nan\c cay, and the Lovell telescope. Parkes PDFB3 observations were retrieved from the CSIRO Data Access Portal. Additional Nan\c{c}ay observations were obtained with the newer Nan\c{c}ay Ultimate Pulsar Processing Instrument (NUPPI; \citealt{Liu14}) at various L-band frequencies with a bandwidth of 512\,MHz. The observations from the Lovell Telescope were obtained with a Digital Filterbank (DFB) since 2010 June and simultaneously with the  Reconfigurable Open Architecture Computing Hardware backend (ROACH; \citealt{Bassa16}) from 2011 April onwards. These instruments observe at centre frequencies of 1524 and 1532\,MHz with bandwidths of 384 and 400\,MHz, respectively. The DFB observations have been previously presented in EPTA DR1 \citep{Desvignes16}. For our analysis, we used the DFB observations for the time period before 2011 April and ROACH observations afterwards, as the latter uses coherent dedispersion to minimise dispersive smearing. 

\begin{figure*}
    \centering
    \includegraphics[width=1\textwidth]{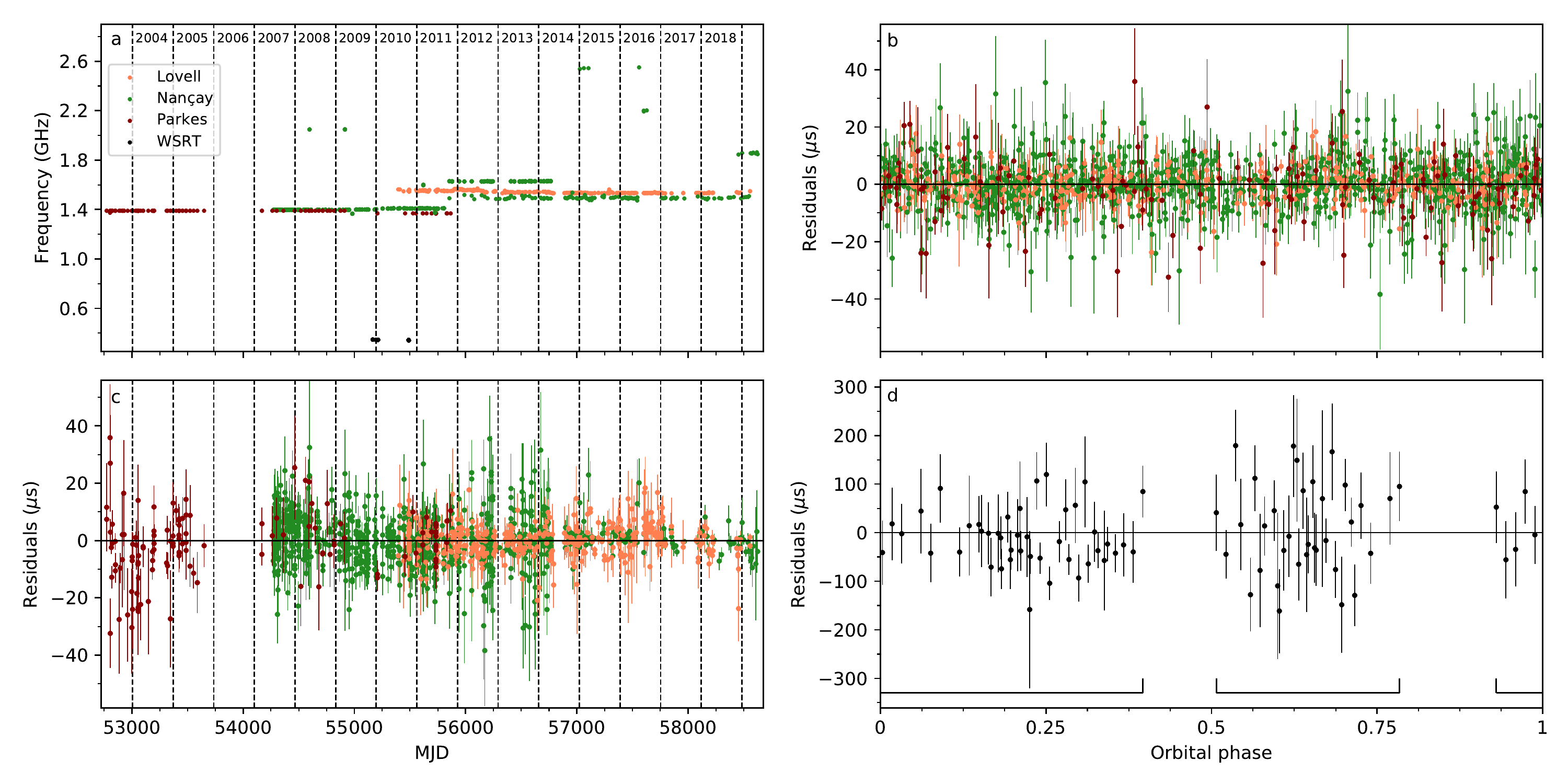}
    \caption{Radio timing observations obtained with the Lovell, Nan\c{c}ay, Parkes, and Westerbork telescopes. \textit{a)}: Timing observations as a function of observing frequency. \textit{b)} and \textit{c)}: The residuals of the timing model as a function of orbital phase and time from observations with Lovell, Nan\c{c}ay, and  Parkes. \textit{d)} The residuals of the timing model as a function of orbital phase from observations with Westerbork. The orbital phases covered by the WSRT observations is indicated with square brackets. Orbital phase $\phi=0$ is defined at the time of ascending node.}
    \label{fig:residuals}   
\end{figure*}

Additionally, six observations from the Westerbork Synthesis Radio Telescope were used to create TOAs and search for radio eclipses at low frequencies. These observations were obtained with the PuMa\,II backend \citep{ksw08}  at 345\,MHz with 70\,MHz of observing bandwidth. An overview of all observations is presented in Table~\ref{tab:obs}. Figure~\ref{fig:residuals}a shows the observing frequency of the observations as a function of time.

\begin{table}[]
    \caption{Overview of the radio observations.}
    \centering
    \footnotesize
    \begin{tabular}{lcrc}
    \hline
        Instrument & \multicolumn{1}{l}{$f_\mathrm{cen}$ (MHz)} & \multicolumn{1}{l}{$N_\mathrm{TOA}$} & \multicolumn{1}{l}{MJD range} \\
    \hline
		Lovell/DFB     & 1524, 1532  & 46 & 55372 - 55669 \\
		\phantom{Lovell/}ROACH   & 1532 & 314 & 55673 - 58557\\[0.3em]
		Nan\c{c}ay/BON     & 1400  & 563 & 54270 - 55806\\ 
		\phantom{Nan\c{c}ay/}BON     & 1600  & 210 & 55619 - 56766\\
		\phantom{Nan\c{c}ay/}BON     & 2000  & 11 & 54596 - 54915\\
		Nan\c{c}ay/NUPPI   & 1484  & 105 & 55854 - 58548 \\
		\phantom{Nan\c{c}ay/}NUPPI   & 1854  & 9 & 58453 - 58626\\
		\phantom{Nan\c{c}ay/}NUPPI   & 2054, 2154, 2539 & 7 & 57026 - 58436\\[0.3em]
        Parkes/AFB     & 1390 & 100 & 52773 - 54911\\
        \phantom{Parkes/}PDFB3   & 1369 & 22 & 55456 - 55866 \\
        \phantom{Parkes/}PDFB4   & 1369 & 4 & 55206 - 55650 \\[0.3em]
		WSRT/PuMa\,II & 345 & 82 & 55164 - 55490 \\
		\hline
    \end{tabular}
    \label{tab:obs}
\end{table}

\subsection{Optical observations}
The faint companion ($V\sim26.7$) to PSR\,J0610$-$2100 was discovered by \citet{Pallanca12} in archival observations obtained with FORS2, the Focal Reducer, and low dispersion spectrograph at the ESO VLT in Chile \citep{Appenzeller98}, in 2004 December and 2005 January. Though only a few $V$ ($9\times1010$\,s) and $R$-band images ($20\times590$\,s) were obtained, these observations showed that the companion was variable in brightness and that the modulation correlated with the orbital period and phase of PSR\,J0610$-$2100.

To improve the light curves, we have obtained follow-up $RI$\footnote{All filters are Bessel $V\!RI$ filters except during this run, when a high throughput $V$-band filter was used.} observations of the companion of PSR\,J0610$-$2100, again with FORS2, in 2010 October, November and December 2010 and 2011 April, with exposure times of $6\times480$\,s and $12\times540$\,s in $R$ and $41\times240$\,s in $I$.

In all observations, FORS2 was used with $2\times2$ binning, providing a pixel scale of $0\farcs25$\,pix$^{-1}$, with the target placed on the master chip. Exposure times were 1010\,s in $V$, 590\,s, 540\,s, or 480\,s in $R$, and 240\,s in $I$. The seeing, as measured from the width of the point spread function (PSF), varied from excellent at $0\farcs27$ to poor at $1\farcs2$, though 80\,per\,cent of the observations had a seeing better than $0\farcs8$. For photometric calibration, short 30\,s $V\!RI$ exposures of the PSR\,J0610$-$2100 field and the Markarian\,A photometric standard field were obtained under photometric conditions during the 2010 October 15th observing run.

\section{Analysis}
\label{analysis}
\subsection{Timing}
We determined TOAs from the folded and dedispersed radio observations of PSR\,J0610$-$2100 using tools from the \textsc{psrchive} software package \citep{Hotan04}. TOAs for the Westerbork observations were fully averaged in frequency and partially averaged in time to about 6\,min per sub-integration. All other observations were fully averaged in time and frequency. After radio frequency interference (RFI) removal, all observations were manually inspected and non-detections and observations still badly affected by RFI were removed. We calculated TOAs by cross correlating the profiles against template profiles. Template profiles were created for each dataset, which we defined as a combination of instrument and centre frequency. Each template consisted of several summed (up to six) von Mises profiles with different phases, widths, and amplitudes to model the averaged profile.

We used the standard timing approach detailed in \citet{Edwards06} where \textsc{tempo2} \citep{Hobbs06} was used to convert the topocentric TOAs to the Solar-system bary-centre with the DE436 solar system ephemeris \citep{Folkner06} and placed them on the Terrestrial Time standard (BIPM2011; \citealt{Petit10}) using Barycentric Coordinate Time (TCB). 

Due to arbitrary phase rotations between templates, as well as cable length differences between instruments, the pulsar timing model includes phase offsets, so-called jumps, between each dataset. We used the recommendation from \citet{Verbiest16} to select the dataset with the lowest value of $\bar{\sigma}_\mathrm{TOA}/\sqrt{N_\mathrm{TOA}}$, where $\bar{\sigma}_\mathrm{TOA}$ is the median TOA uncertainty and $N_\mathrm{TOA}$ the number of TOAs in that dataset, as the reference dataset to reference jumps to. This is the NUPPI dataset at 1484\,MHz. Due to the simultaneity of some of the Lovell DFB and ROACH observations, phase offsets between these datasets were determined directly and not included as free parameters in the timing model. Afterwards, TOAs from Lovell DFB observations that were obtained simultaneously with ROACH observations were removed. Phase offsets between the previously published Parkes PDFB3 TOAs and the additional PDFB3 TOAs from folded profiles were determined in a similar fashion, as there was an overlap between these datasets. After obtaining a phase offset, the TOAs from the folded profiles that coincided with the previously published PDFB3 TOAs were removed.

We started with the EPTA DR1 pulsar timing model of PSR\,J0610$-$2100 by \citet{Desvignes16}, which includes proper motion, a polynomial model for variations in dispersion measure ($\mathrm{DM}$), and models the low-eccentricity binary orbit of PSR\,J0610$-$2100 with the ELL1 binary model \citep{Lange01}. The ELL1 model removes the strong covariance between eccentricity $e$ and longitude of periastron $\omega$ for small eccentricities by fitting the Laplace-Lagrange parameters $\eta=e\sin\omega$ and $\kappa=e\cos\omega$ instead. TOAs from the Lovell, Nan\c{c}ay, and Parkes telescopes were included in the fit, while the Westerbork TOAs were only used to assess the presence of a radio eclipse or orbital phase dependent DM variations (see below). After an initial fit of the timing model using all datasets, we applied an iterative procedure where the profiles were refolded with the new model to create improved templates with which TOAs were redetermined. To minimise the impact of outliers on the fit, TOAs for which no profiles were available and which resulted in outlying residuals ($>3\sigma$)  were removed. Additionally, any TOAs with uncertainties exceeding 20\,ms were also removed.

After three such iterations, we obtained the timing model given in Table~\ref{tab:par1}. The residuals for this model are displayed as a function of orbital phase and time in Fig.~\ref{fig:residuals}b and \ref{fig:residuals}c. To allow for comparison with the \citet{Desvignes16} timing model, we fitted a pulsar timing model using the same reference epoch and time of ascending node. Furthermore, we fitted a timing model where these are referenced to the centre of the timespan covered by the TOAs, to minimise covariances between fitted parameters. For the latter model, we included orbital period derivatives to fit for variations in the orbital period ($P_\mathrm{b}$).

\subsection{Orbital period variations}\label{ssec:orbvars}
Since high precision pulsar timing of other black widow systems show variations of the orbital parameters (e.g.\ \citealt{Shaifullah16}), we checked for variability by computing pulsar timing models for segments of the full timing baselines of 16\,yr to constrain the presence of similar variations in PSR\,J0610$-$2100. Three different segment lengths of 180, 360, and 720\,days were used, with 15, 30, and 60-day overlaps, respectively. For each segment we used \textsc{tempo2} to propagate the timing model from Table~\ref{tab:par1} to a reference epoch at the centre of the segment, a step which adjusts the position, spin frequency, and $\mathrm{DM}$ for the effects of proper motion and spindown. The time of ascending node $T_\mathrm{ASC}$ was also adjusted to the centre of each segment. Next, we refitted the orbital parameters (orbital period $P_\mathrm{b}$, $T_\mathrm{ASC}$, the projected semi-major axis $x$, $\eta$, and $\kappa$) for each segment, keeping all other parameters fixed. These fits did not include $\dot{P}_\mathrm{b}$, which was set to zero.

Figure~\ref{fig:orbvars} shows the resulting variations in the orbital parameters for the different segment lengths compared to the parameters of the timing model in Table~\ref{tab:par1}. These are discussed in \S\,\ref{results}. 

\subsection{Optical photometry}
\label{s:optical}
The FORS2 observations were corrected for bias using bias frames obtained during daytime and flat-fielded using twilight sky flats. To align the images all were registered using integer pixel offsets. A $2\farcm1\times2\farcm1$ subsection of these images was analysed with DAOPHOT\,II \citep{Stetson87}, where instrumental magnitudes were determined using PSF fitting. Offsets in instrumental magnitude between the 30\,s exposure $V\!RI$ images and images obtained using the same filter were measured to place the magnitudes on the same zeropoint. The instrumental magnitudes were subsequently calibrated using 70 photometric standards from the Markarian\,A field using calibrated Johnson-Cousins $V\!RI$ magnitudes by \citet{Stetson00}. We fitted for zeropoint and colour terms, but used tabulated ESO extinction coefficients of 0.113, 0.109, and 0.087 mag\,airmass$^{-1}$ for $V\!RI$, respectively. The rms residuals of these calibrations were 0.09\,mag in $V$, 0.05\,mag in $R$ and 0.11\,mag in $I$.

\subsection{Optical modelling}
\label{s:icarus}
We modelled the optical photometry using the \texttt{Icarus} binary light curve modelling software \citep{Icarus}\footnote{\url{https://github.com/bretonr/Icarus}}. \texttt{Icarus} generates a model star with a tessellated surface accounting for the tidal deformation in the binary potential. It then computes photometric fluxes in each filter band, at each orbital phase, for each surface element according to the local surface temperature, gravity, and viewing angle, and sums the fluxes from all surface elements to produce a model light curve. Here, we computed these fluxes from the model spectra of the G\"{o}ttingen Spectral Library\footnote{\url{http://phoenix.astro.physik.uni-goettingen.de/}}
\citep{Husser2013+Atmos}.

We used the \texttt{Multinest} nested sampling algorithm \citep{MultiNest} to explore the parameter space and estimate posterior distributions for model parameters. For exposures in which the optical counterpart to PSR~J0610$-$2100 was not detected, we assumed a flux of zero and an uncertainty that was one third of the image's 3$\sigma$ detection threshold, and included these points in our fitting to ensure the model predicted fluxes below the detection limits around the minimum.
We allow for 0.1\,mag offsets in the flux calibrations in each band (guided by the residuals quoted in Section~\ref{s:optical}), which accounts for systematic uncertainties in both the photometric calibration, and in the continuum level of the model spectra from which \texttt{Icarus} computes fluxes. 

The parameters of interest in our models were: the binary inclination angle, $i$, for which we assumed a prior probability of $\sin(i)$ between $0^\circ$ (a face-on orbit) and $90^\circ$ (edge-on); the Roche-lobe filling factor, $f_{\rm RL}$, defined as the ratio between the stellar and Roche-lobe radii in the direction from the companion's centre-of-mass to the L1 Lagrange point; the base temperature, $T_{\rm base}$, of the star before irradiation and gravity darkening are applied; the irradiating temperature $T_{\rm irr}$ (see below); and the distance, $d$. For the distance, we assumed a prior that combines a model of the millisecond pulsar density within the Galaxy \citep{Levin2013+GalMSPs} as a function of the distance along the line of sight, multiplied by a log-Gaussian distribution centred on the DM distance of $3.26$\,kpc (see \S\,\ref{results}), with $1\sigma$ fractional uncertainties of $\pm45\%$. 

In addition, we fitted for the extinction, $E(B-V)$, and pulsar mass $M_{\rm psr}$, neither of which are constrained by the data, but which were included here as `nuisance' parameters and marginalised over to account for their effects on our parameter uncertainties. For the extinction, we adopted a Gaussian prior of $E(B-V)=0.05\pm0.02$, based on the estimate from \citet{Bayestar19} for the extinction towards PSR~J0610$-$2100, truncated such that $E(B-V) > 0$. For the pulsar mass, we assumed a uniform prior between 1.0\,M$_{\odot} < M_{\rm psr} < 2.5$\,M$_{\odot}$, encompassing the observed neutron star mass range \citep{of16}. To account for gravity darkening, the surface temperature was modified according to the local surface gravity $g$, by $T\propto g^{\beta}$, with a fixed index of $\beta = 0.08$, which assumes that the companion star has a convective envelope. 

We also derived additional parameters using the sampled values of the optical model parameters, along with the parameters obtained from radio timing (see Table \ref{tab:par1}): the pulsar's semi-major axis, $x$; orbital period, $P_{\rm b}$; spin period, $P$; and intrinsic spin-down rate, $\dot{P}_{\rm int}$. The companion mass $M_{\rm comp}$ and binary mass ratio, $q \equiv M_{\rm psr}/M_{\rm comp}$ were derived from $M_{\rm psr}$, $i$, and the binary mass function $f(M) = M_{\rm comp}^3 \sin^3 i / (M_{\rm comp} + M_{\rm psr})^2 = 4 \pi^2 x^3 / (G P_{\rm b} ^2) = 5.2\times10^{-6}$\, M$_{\odot}$. The heating efficiency, $\epsilon \equiv L_{\rm irr}/\dot{E}_{\rm int}(d)$ \citep{Breton2013} is the ratio between the heating luminosity, $L_{\rm irr} = 4 \pi A^2 \sigma T^4_{\rm irr}$ where $A = x(1+q)/\sin i$ is the orbital separation, and the pulsar's intrinsic spin-down power budget, $\dot{E}_{\rm int}(d) = 4 \pi^2 I \dot{P}_{\rm int}(d) / P^3$ (where the distance dependence is due to the Shklovskii correction described below, and we assumed that the pulsar's moment of inertia $I=10^{45}$\,g\,cm$^2$). In our model, we assumed that the pulsar acts as a point source of an isotropic heating flux and that this flux raises the surface temperature when it impinges on the companion star. The efficiency $\epsilon$ therefore incorporates unknown beaming and stellar albedo factors. While $\epsilon > 1$ is possible (e.g.\ if the heating flux is strongly beamed towards the companion), we excluded models requiring $\epsilon > 200$, effectively ruling out large distances above which $\dot{E}_{\rm int}$ becomes very low or negative.

\begin{table}
    \footnotesize
    \caption{Pulsar timing model for PSR J0610$-$2100. Figures in parentheses are  the nominal 1$\sigma$ \textsc{tempo2} uncertainties in the least-significant digits quoted, which have been multiplied by the square root of the reduced $\chi^2$.}
    \begin{tabular}{ll}
        \hline\hline
        \multicolumn{2}{c}{Dataset and assumptions} \\
        \hline
        MJD range & 52772.1---58626.6 \\
        Reference epoch (MJD) & 55699 \\
        Data span (yr) & 16.03 \\
        Number of TOAs & 1391 \\
        RMS timing residual ($\upmu$s) & 3.6 \\
        Weighted fit &  Y \\
        Reduced $\chi^2$ value  & 1.6 \\
        Clock correction procedure & TT(BIPM2011) \\
        Solar system ephemeris model & DE436 \\
        Binary model & ELL1 \\
        \hline
        \multicolumn{2}{c}{Measured Quantities} \\
        \hline
        Right ascension, $\alpha_\mathrm{J2000}$ & $06^\mathrm{h}10^\mathrm{m}13\fs596735(7)$ \\
        Declination, $\delta_\mathrm{J2000}$ & $-21\degr00\arcmin27\farcs89913(11)$ \\
        Pulse frequency, $\nu$ (s$^{-1}$) & $258.978475098116(6)$ \\
        First derivative of pulse frequency, $\dot{\nu}$ (s$^{-2}$) & $-8.26744(7)\times 10^{-16}$ \\
        Dispersion measure, $\mathrm{DM}$ (cm$^{-3}$\,pc) & $60.6722(11)$ \\
        First derivative of $\mathrm{DM}$, $\dot{\mathrm{DM}}$ (cm$^{-3}$\,pc\,yr$^{-1}$) & $-0.0026(4)$ \\
        Proper motion in $\alpha_\mathrm{J2000}$, $\mu_{\alpha} \cos \delta$ (mas\,yr$^{-1}$) & $9.11(3)$ \\
        Proper motion in $\delta_\mathrm{J2000}$, $\mu_{\delta}$ (mas\,yr$^{-1}$) & $16.45(3)$ \\
        Orbital period, $P_\mathrm{b}$ (d) & $0.28601600633(3)$ \\
        First derivative of orbital period, $\dot{P}_\mathrm{b}$ (s\,s$^{-1}$) & $-7(3)\times 10^{-14}$ \\
        Projected semi-major axis of orbit, $x$ (lt-s) & $0.07348914(14)$ \\
        Time of ascending node passage $T_\mathrm{asc}$ (MJD) & $55699.0070228(2)$ \\
        $\eta=e\sin\omega$ & $2.3(5)\times 10^{-5}$ \\
        $\kappa=e\cos\omega$ & $4(4)\times 10^{-6}$ \\
        \hline
         \multicolumn{2}{c}{Derived Quantities} \\
        \hline
        First derivative of pulse period, $\dot{P}$ (s\,s$^{-1}$) & $1.232661(10)\times 10^{-20}$\\
        Transverse proper motion, $\mu$ (mas\,yr$^{-1}$) & $18.81(3)$ \\
    \end{tabular}
    \label{tab:par1}
\end{table}

\begin{figure*}
   \centering
   \includegraphics[width=0.9\textwidth]{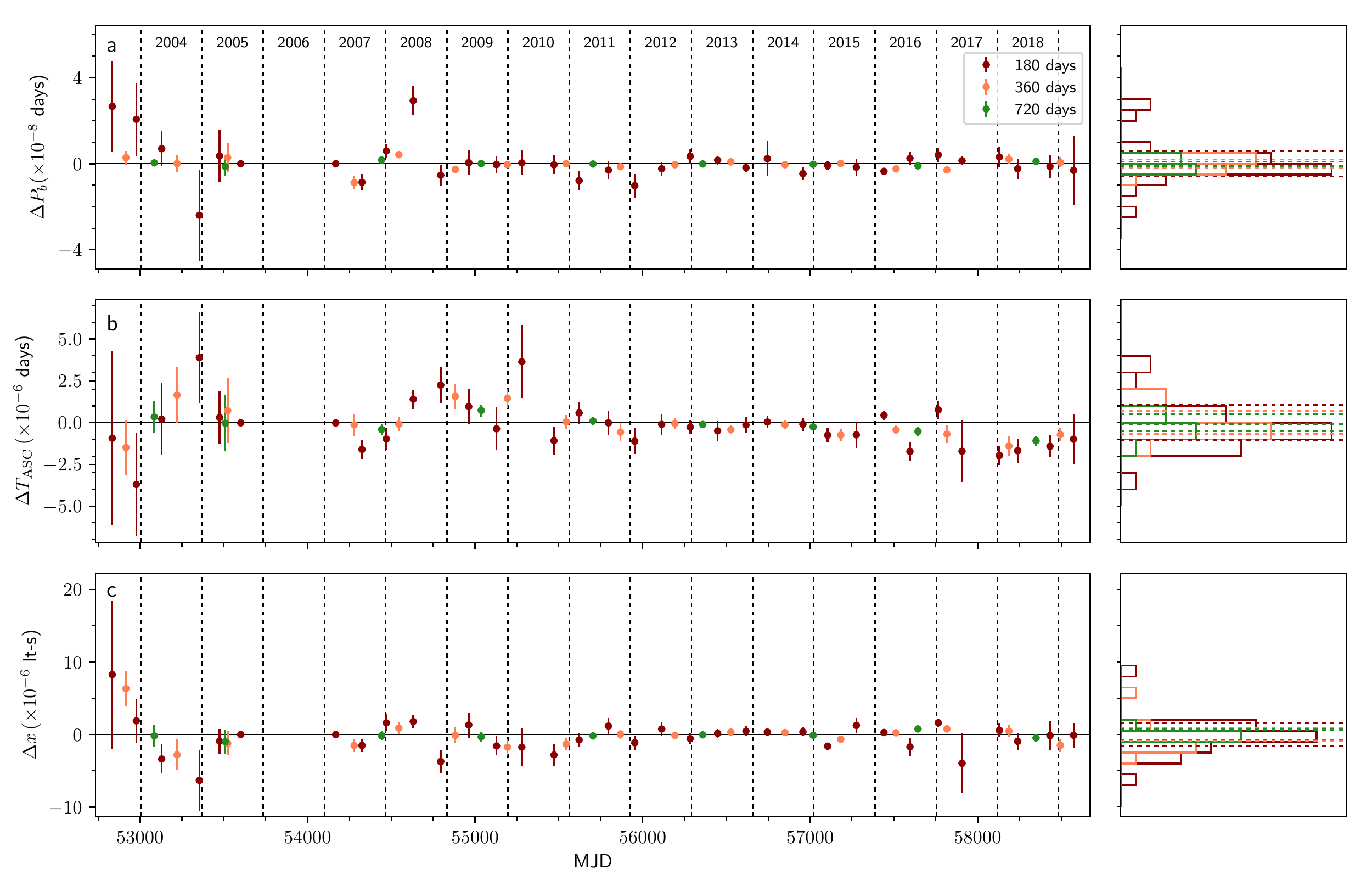}
   \caption{Constraints on orbital variations. Variations in the orbital period $\Delta P_\mathrm{b}$ (panel a), the epoch of the ascending node $\Delta T_\mathrm{ASC}$ (panel b), and the projected semi-major axis of the orbit $\Delta x$ (panel c) are plotted against time for 180-, 360-, and 720-day segments. Here $\Delta P_\mathrm{b}=0$ refers to the orbital period in Table \ref{tab:par1} found for the full span and similarly for the other parameters. The right-hand panels show distributions of the variations for each segment length. The dotted lines denote the mean of the uncertainty on the measurements. The uncertainties are generally of the same order of magnitude as the standard deviations on the variations and the measurements do not show any clear trends.}
   \label{fig:orbvars}
\end{figure*}

\section{Results}  
\label{results}
\subsection{Timing}
The pulsar timing dataset presented here has a timespan of 16\,years, doubling that of \citet{Desvignes16}. In general, the pulsar timing solution derived from our datasets is consistent with that found by \citet{Desvignes16}. The few parameters that differ at the $3\sigma$ level are the astrometric parameters and the dispersion model. We attribute these differences to our longer dataset. The uncertainties on the rotational, astrometric, and binary parameters have decreased significantly. We note that the Laplace-Lagrange parameters yield an eccentricity of $e=2.3(4)\times 10^{-5}$ with longitude of periastron $\omega=79(10)\degr$, also consistent with the values obtained by \citet{Desvignes16}.

Due to the large proper motion of PSR\,J0610$-$2100 of $\mu=18.81(3)$\,mas\,yr$^{-1}$, the pulsar is subject to an apparent acceleration, the Shklovskii effect \citep{Shklovskii70} of $c\dot{P}_\mathrm{shk}/P=\mu^2 d$ which offsets the observed spin period derivative $\dot{P}_\mathrm{obs}=1.232661(10)\times 10^{-20}$\,s\,s$^{-1}$. We can use this apparent acceleration to set an upper limit to the distance of the pulsar of $d<3.73$\,kpc, above which the intrinsic spin period derivative $\dot{P}_\mathrm{int}=\dot{P}_\mathrm{obs}-\dot{P}_\mathrm{shk}$ would be negative and hence unphysical. This is consistent with the DM derived distance estimates of 3.54\,kpc from the NE2001 model \citep{Cordes02} and 3.26\,kpc from the YMW16 model \citep{Yao17}, yielding $\dot{P}_\mathrm{shk}=1.17\times10^{-20}$\,s\,s$^{-1}$ and $1.08\times10^{-20}$\,s\,s$^{-1}$, respectively. Other apparent accelerations impacting the spin period derivative are caused by differential Galactic rotation, introducing $\dot{P}_\mathrm{dgr}$, and acceleration towards to the Galactic disk $\dot{P}_\mathrm{kz}$ \citep{Damour91, Nice95}. For distances up to $d=3.73$\,kpc, the sum of these effects ranges from $-1.8\times10^{-22}$ to $-1.2\times10^{-22}$\,s\,s$^{-1}$, two orders of magnitude smaller than that caused by the Shklovskii effect. 

Similarly, these effects also impact the observed orbital period, with the Shklovskii effect dominating. For the above distance upper limit, we obtained $\dot{P}_\mathrm{b, shk} < 7.91\times 10^{-14}$\,s\,s$^{-1}$. The observed value of the orbital period derivative $\dot{P}_\mathrm{b,obs}=-7(3)\times10^{-14}$\,s\,s$^{-1}$ in our timing solution is not significant. In none of the models including higher order derivatives of $P_\mathrm{b}$ were any of the derivatives significant ($>3\sigma$). From refitting the orbital parameters for smaller time segments of the full time span, we found that the scatter in the measurements of $P_\mathrm{b}$, $T_\mathrm{ASC}$, and $x$ are of the same order of magnitude as their uncertainties and, except for $T_\mathrm{ASC}$, do not show any clear trends or periodic structure (Fig.\,\ref{fig:orbvars}). There is a possible weak trend seen in $T_\mathrm{ASC}$ during 2017 and 2018, but its significance is low ($\sim3\sigma$). To quantify the variations, we obtained the standard deviation of the normalised variations on the shortest timescales as $\sigma_{\Delta P_\mathrm{b}}/P_\mathrm{b} = 3.3\times10^{-8}$, $\sigma_{\Delta T_\mathrm{asc}}/P_\mathrm{b} = 5.1\times10^{-6}$, and $\sigma_{\Delta x}/x = 3.2\times10^{-5}$. Hence, we conclude that no significant orbital variations of $P_\mathrm{b}$ and $x$ are present in our observations of PSR\,J0610$-$2100 while slight variations in $T_\mathrm{ASC}$ may be present. 
    
Previously, no eclipses of the radio signal have been detected in observations of PSR\,J0610$-$2100 between frequencies of 1374 and 2048\,MHz. Confirming previous findings, our observations show no evidence of radio eclipses near superior conjunction at orbital phase $\phi=0.25$ (Fig.\,\ref{fig:residuals}d) down to observing frequencies of 310\,MHz.  The residuals near superior conjunction do not show additional dispersive delays due to a ionised material evaporating from the pulsar companion. The rms in the residuals between the orbital phases of $\phi=0.2$ and 0.3 sets a limit on additional DM delays of 0.02\,pc\,cm$^{-3}$ at 1532\,MHz and 0.005\,pc\,cm$^{-3}$ at 345\,MHz.

\subsection{Optical modelling}
\label{s:icarus_results}
\begin{figure}
  \centering
  \includegraphics[width=\columnwidth]{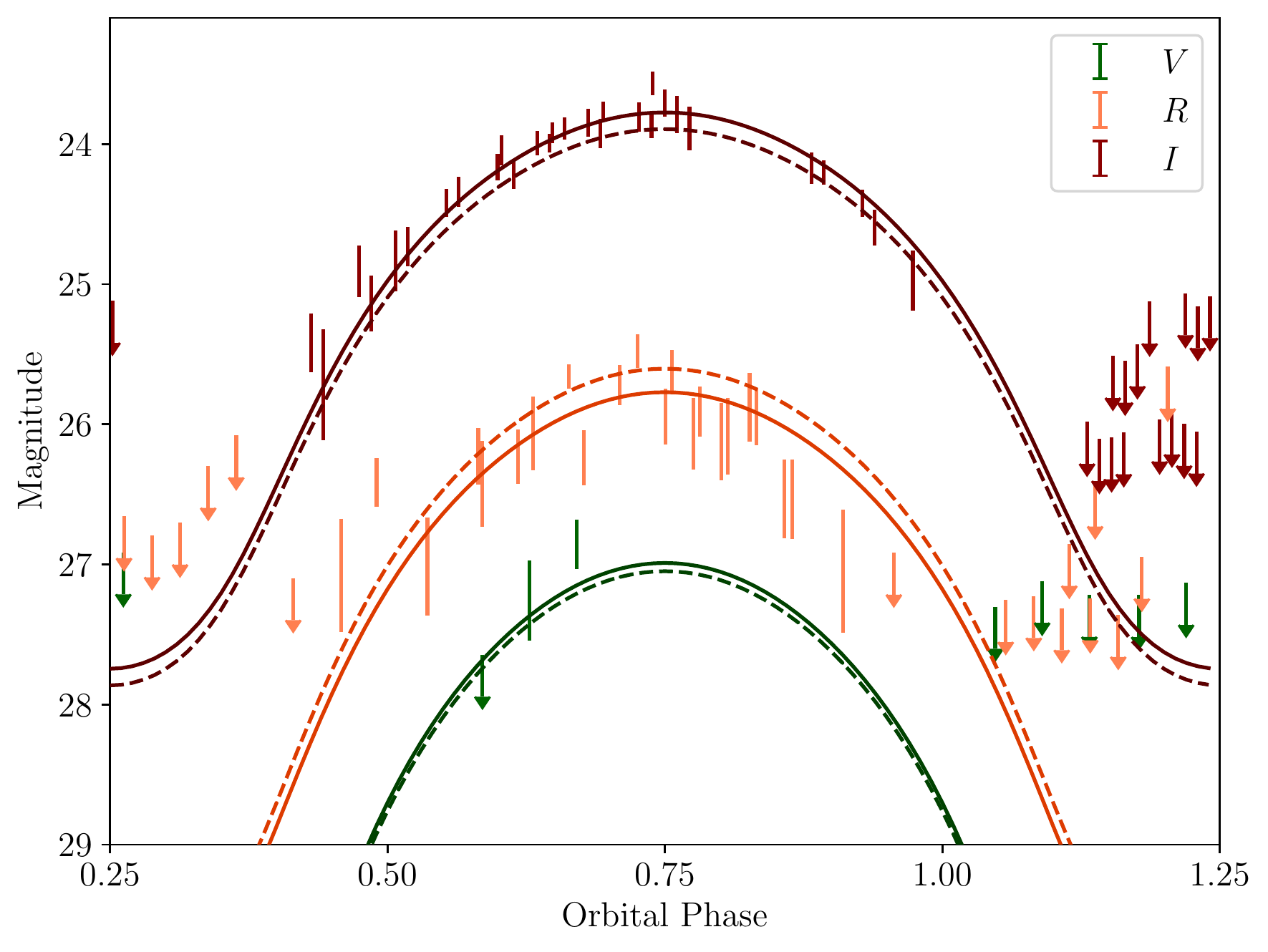}
  \caption{Optical light curve and \texttt{Icarus} model for the companion star of PSR~J0610$-$2100. Vertical lines show the measured magnitudes and their $1\sigma$ uncertainties. Where the counterpart is not detected in an image, we plot $2\sigma$ upper limits instead, denoted by downward-pointing arrows. Dashed curves show the flux in each band as predicted by our best-fitting \texttt{Icarus} model, solid curves show the same model after allowing for small ($0.1$\,mag) offsets between the photometric calibration and the model normalisation. An orbital phase of $\phi=0$ corresponds to the pulsar's ascending node; the companion star's superior conjunction is at phase $\phi=0.75$. \label{f:icarus_model}}
\end{figure}

\begin{figure*}
  \centering
  \includegraphics[width=\linewidth]{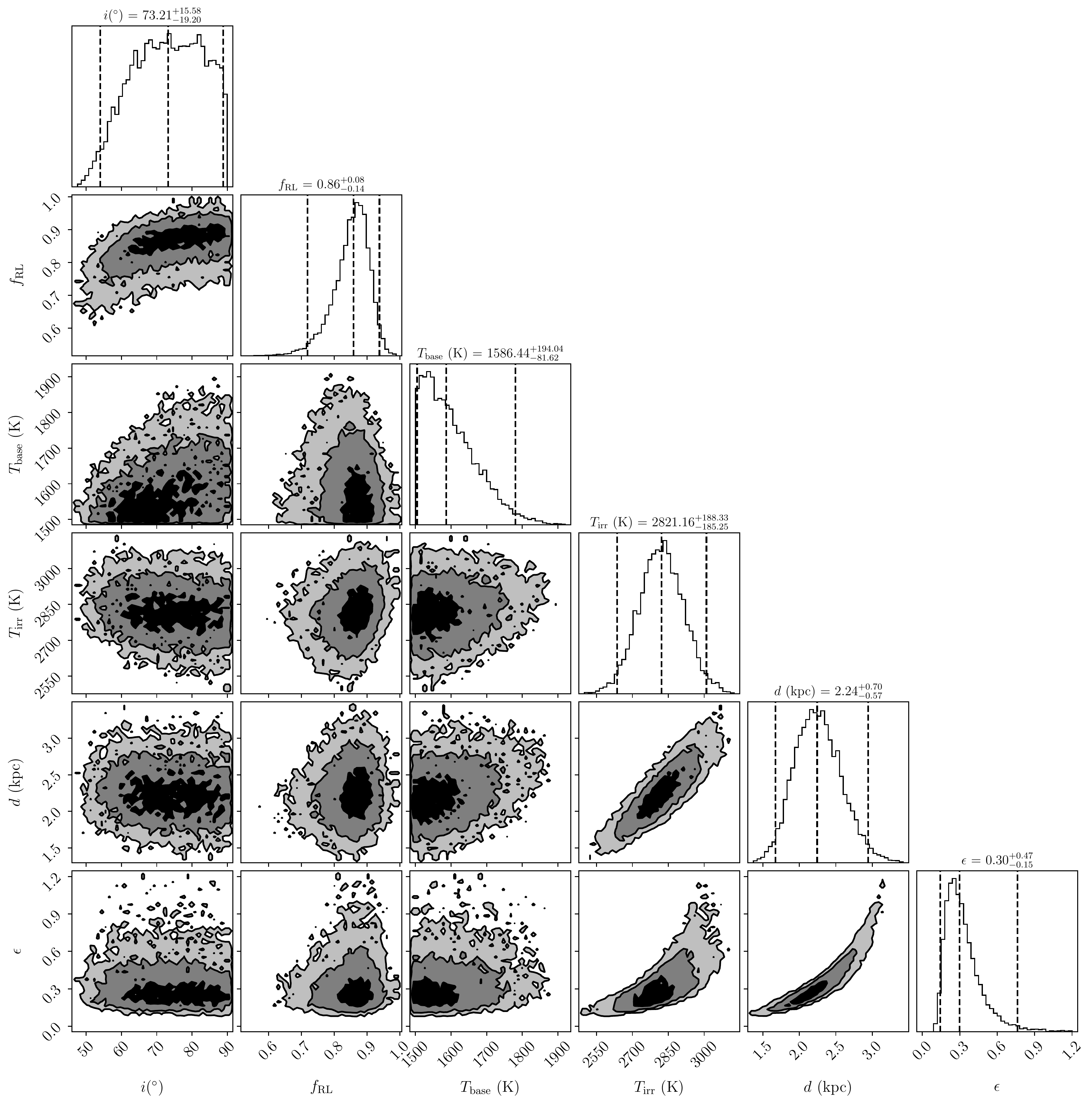}
  \caption{Posterior distributions for \texttt{Icarus} model parameters. Contours on the 2-d conditional distributions are at the $1\sigma$, $2\sigma$ and $3\sigma$ levels. 1-d marginalised distributions are shown on the diagonal. The final parameter, $\epsilon$, is derived from $i$, $T_{\rm irr}$, $d$ (which enters via the Shklovskii corrections to $\dot{E}_\mathrm{int}$) and the pulsar mass (not shown), which is not constrained by the data, but which we marginalised over as a nuisance parameter.
    \label{f:icarus_params}}
\end{figure*}

The best-fitting model of the optical light curve for the companion star is shown in Figure~\ref{f:icarus_model} with posterior distributions for the fit and derived parameters shown in Figure~\ref{f:icarus_params}. The best-fitting model has a reduced chi-square of $\chi^2 / n = 98.3 / 82 = 1.20$. All quoted uncertainties here are 95\% confidence intervals from the posterior distributions estimated by MultiNest.

From our Icarus models, we find that the base temperature $T_{\rm base} < 1800$\,K is very low, with the samples converging towards the lower limit allowed in our modelling of 1500\,K. This limit is quite far below the lowest temperature (2300\,K) covered by the spectral library used by \texttt{Icarus} to compute fluxes. Extrapolating to temperatures far below the range of the spectral libary is problematic, as unmodelled effects such as dust settling start to become important when simulating low-temperature spectra \citep{Husser2013+Atmos}. While this unreliable extrapolation must be treated with caution, this will not badly affect our fits, as the model light curves are dominated by hotter surface elements on the irradiated side of the star. Nevertheless, the inferred temperature values far below the 2300\,K limit of the spectral library cannot be trusted. We therefore conclude only that the base temperature must be well below the lower limit of the spectral library. 

We find also that the irradiation temperature $T_{\rm irr} = 2820 \pm 190$\,K is unusually low in this system, with $T_{\rm irr} > 4000$\,K being more typical for black widow companions \citep[e.g.][and see Figure \ref{f:irradiation}]{Breton2013}. However, this weak irradiation seems to be consistent with the pulsar's low intrinsic spin-down power ($1.9\times10^{33}$\,erg\,s$^{-1} < \dot{E}_{\rm int} < 4.8\times10^{33}$\,erg\,s$^{-1}$), as the heating efficiency $0.15 < \epsilon < 0.77$ is typical for black widow systems. Owing to the low base temperature, this weak irradiation is still sufficient to produce a large relative temperature difference between the heated and unheated sides and hence the observed 4-mag variation in the $I$-band flux across the orbit.

The mechanism through which pulsars heat their companion stars remains unclear, but one explanation is that the pulsar's gamma-ray flux is deposited in the upper stellar atmosphere of the companion. The heating efficiency can therefore be compared to the efficiency through which the pulsar converts its spin-down power into gamma-ray emission, $\eta_{\gamma} \equiv 4 \pi F_{\gamma} d^2 / \dot{E}_\mathrm{int}(d)$, where $F_{\gamma}$ is the observed gamma-ray flux and $d$ is the best fitting distance of 2.24\,kpc. The DR2 \citep{4FGLDR2} iteration of the Fermi-LAT Fourth Source Catalogue \citep{4FGL} gives the gamma-ray flux of PSR~J0610$-$2100 above 100\,MeV as $F_{\gamma} = (6.8 \pm 0.5) \times10^{-12}$~erg~cm$^{-2}$~s$^{-1}$, which corresponds to $0.5 < \eta_{\gamma} < 3.7$ (where efficiencies above unity suggest non-isotropic emission preferentially beamed towards the observer). The ratio between $\epsilon$ and $\eta_{\gamma}$ is better constrained than either individual efficiency (as the distance dependence in $\dot{E}_{\rm int}$ cancels out), which we find to be $0.14 < \epsilon/\eta_{\gamma} < 0.34$, suggesting that the pulsar's gamma-ray flux has sufficient power to explain the irradiation in this system. However, we note that in the overall population of black widows, these efficiencies are not well correlated with one another \citep{Draghis2019}. 

We find that the Roche-lobe filling factor is preferred to be below unity ($0.72 < f_{\rm RL} < 0.94$). We also performed the fitting with a fixed $f_{\rm RL} = 1.0$, which gave parameter estimates that were consistent with those found when $f_{\rm RL}$ was free, albeit with slightly higher $i$ and $T_{\rm irr}$, but with slightly worse chi-square $\Delta \chi^2 = 3.8$ and log-evidence $\Delta \log Z = 2.6$\footnote{The evidence is the integral of the likelihood over the prior distribution.}.

The observed light curve in Figure~\ref{f:icarus_model} has hints of asymmetry, with data points after the conjunction at orbital phase $\phi=0.75$ being systematically lower than those at the equivalent pre-conjunction phase. Such asymmetries in black-widow light curves can be caused by various effects: reprocessing of heating flux by an asymmetric intra-binary shock \citep{Romani2016+IBS}; channelling of charged particles onto the magnetic poles of the companion star \citep{Sanchez2017+Bduct}; or convective winds on the stellar surface \citep{Kandel2020+conv,Voisin2020+redist}. We therefore repeated our modelling with the addition of a hot spot (whose size, temperature, and location were new free parameters) on the trailing face of the companion star to mimic these effects. This model gave a better chi-squared, thanks to the additional degrees of freedom,  with a slightly lower range of filling factors ($f_{\rm RL} = 0.78^{+0.1}_{-0.13}$) and slightly higher inclination ($i > 62\degr$), but all parameters were consistent within the uncertainties of the simpler symmetric-heating model.

\section{Discussion and conclusions}
\label{discussion}
The radio timing measurements of PSR\,J0610$-$2100 show no evidence of radio eclipses down to low frequencies ($>310$\,MHz) or significant orbital period variations over the 16\,yr timing baseline. Modelling of the optical light curve of PSR\,J0610$-$2100's  low-mass companion confirms the irradiation of the companion by the pulsar, as previously reported by \citet{Pallanca12}, and shows that the companion is likely not filling its Roche lobe, while having a high inclination ($54\degr<i<89\degr$). We found a surprisingly low temperature for the irradiated hemisphere of the companion. 

\begin{figure}
  \centering
  \includegraphics[width=\columnwidth]{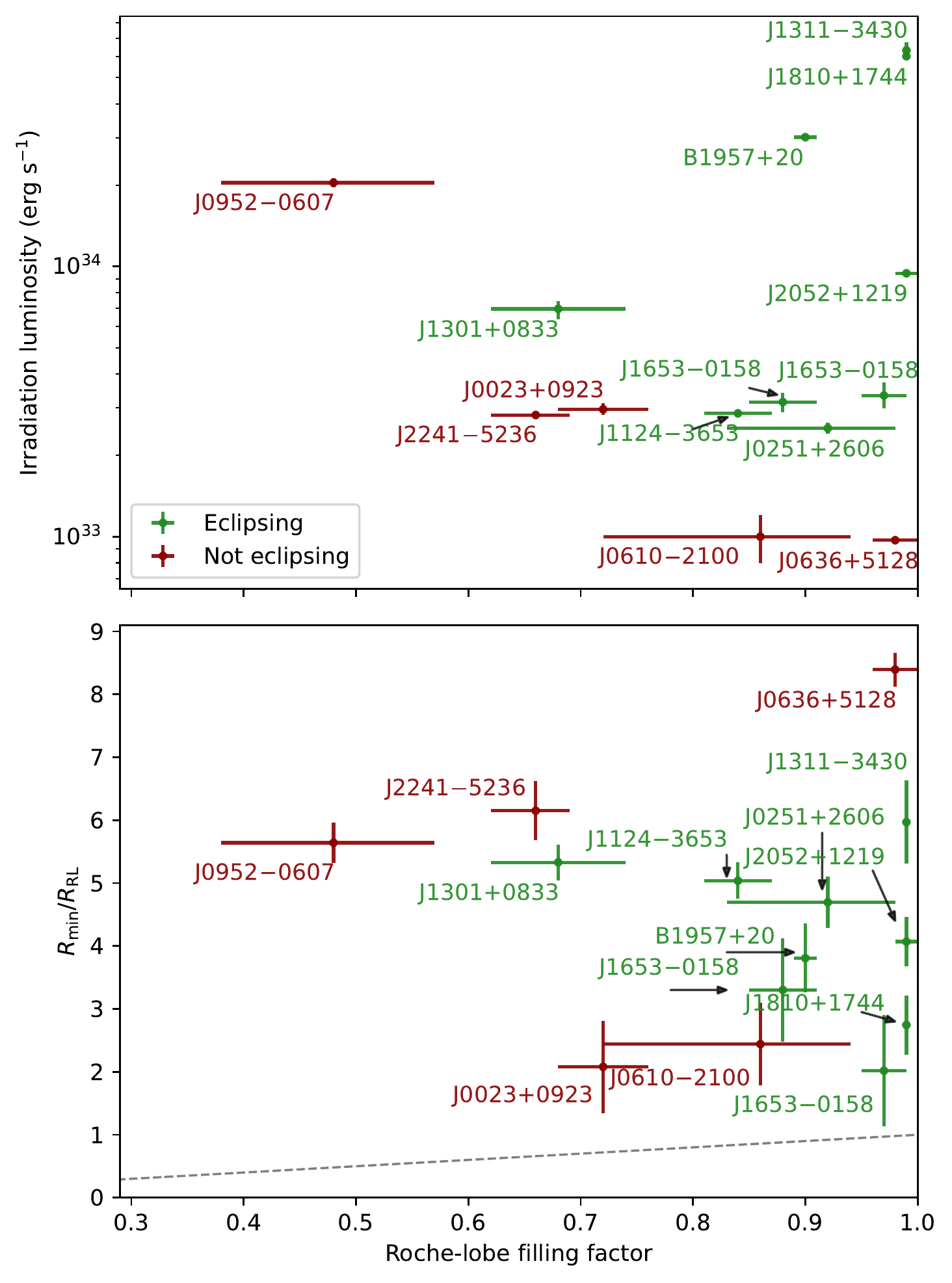}
  \caption{\label{f:irradiation} Companion properties and presence of radio eclipses of black widow systems for which the optical light curves of irradiated companions have been modelled with \textsc{Icarus}. Higher irradiation luminosities and Roche-lobe filling factors are preferred for eclipsing black widow systems compared to non-eclipsing systems. The dashed line in the bottom panel denotes the companion radius for which physical eclipses would occur. The properties of PSR\,J1653$-$0158 are based on two different light curve models (see \citealt{Nieder2020} for details). }
\end{figure}

While eclipses of the pulsed radio emission are seen in several black widow systems, the absence of radio eclipses in PSR\,J0610$-$2100 is not unique. We compared companion properties between all black widow systems modelled with \textsc{Icarus} to investigate the relation between irradiation, Roche-lobe properties, and the presence of radio eclipses. \citet{Draghis2019} modelled optical light-curves of irradiated black widow companions with \textsc{Icarus} for nine systems, of which no radio eclipses are observed in four of them (PSRs\,J0023+0923; \citealt{Bak20}, J0636+5128; \citealt{Stovall14}, J0952$-$0607; \citealt{Bassa17}, and J2241$-$5236; \citealt{Keith2011}).  Radio eclipses have been observed in PSRs\,J0251+2606, J1124$-$3653, J1301+0833, B1957+20, and J2052+1219 (\citealt{Deneva2021,Fruchter88}, J.\ Deneva and P.\ Ray, priv.\ comm.). Furthermore, light-curve modelling has also been performed for PSR\,J1810+1744 \citep{Romani2021} and PSR\,J1653$-$0158 \citep{Nieder2020}, where the PSR\,J1810+1744 shows radio eclipses \citep{Polzin2018}, while PSR\,J1653$-$0158 remains undetected in radio, which we attribute to eclipses of the radio emission over large fractions of the orbit \citep{Nieder2020}. 

Radio eclipses or orbitally modulated DM variations detected in the pulsar observations are caused by (ionised) material in the evaporative wind from the companion passing the line of sight. Conversely, the absence of eclipses can be caused by either the absence of ionised material, or by the material missing the line of sight. The mass lost in the evaporative wind is predicted to be higher for systems filling their Roche lobe, compared to detached systems ($f_\mathrm{RL}<1$) and dependent on the irradiation luminosity \citep{Ginzburg+2020}. This is qualitatively confirmed in Fig.\,\ref{f:irradiation}, which collates the companion properties and highlights the presence or absence of radio eclipses; the black widow systems showing radio eclipses tend to have a higher Roche-lobe filling factor and irradiation luminosity compared to the non-eclipsing systems. Material missing the line of sight is predominantly dependent on the orbital inclination and a statistical analysis of a larger sample of black widow pulsars indicates that pulsar mass functions for eclipsing black widow systems are higher than non-eclipsing systems, indicating that the former have higher inclinations \citep{Guillemot19}. We quantified this in the lower panel in Fig.\,\ref{f:irradiation} where the minimum distance $R_\mathrm{min}$ between the companion and the line of sight is compared to the Roche-lobe radius $R_\mathrm{RL}$. The distance $R_\mathrm{min}$ is a strong function of inclination through the orbital projection, which sets $R_\mathrm{min} = x(1+q)/\tan{i}$, as well as mass ratio $q$ constrained by the inclination and the mass function (we assumed a pulsar mass of 1.5\,M$_\odot$). Furthermore, the Roche-lobe radius depends on inclination through the mass ratio. For $R_\mathrm{min}/R_\mathrm{RL}<1$ and $(R_\mathrm{min}/R_\mathrm{RL})/f_\mathrm{RL}<1$ the Roche lobe and actual companion radius would actually pass the line of sight. We find that for those systems with radio eclipses, the eclipses are caused by material that extends multiple Roche-lobe radii away from the companion. PSR\,J0610$-$2100 is one of the few non-eclipsing systems where the companion passes within a few Roche-lobe radii from the line of sight. Hence, it is tempting to speculate that the low spin-down of PSR\,J0610$-$2100 and the correspondingly low temperature of the irradiated side of the companion, resulting in less mass lost in the evaporative wind, may explain the absence of radio eclipses observed in this system. 

Where the original black widow pulsars show large orbital variations leading to $\Delta T_\mathrm{ASC}/P_\mathrm{b}\approx6\times10^{-4}$, $\Delta P_\mathrm{b}/P_\mathrm{b}\approx1\times10^{-7}$, and $\Delta x/x\approx9\times10^{-4}$ in the case of PSR\,J2051$-$0827 \citep{Shaifullah16} and $\Delta P_\mathrm{b}/P_\mathrm{b}\approx1.6\times10^{-7}$ for PSR\,B1957+20 \citep{Arzoumanian94}, these variations were not detected in $P_\mathrm{b}$ and $x$ for PSR\,J0610$-$2100, with limits of $|\Delta P_\mathrm{b}|/P_\mathrm{b}<3.3\times10^{-8}$ and $|\Delta x|/x < 3.2\times10^{-5}$. There is a possible weak trend ($\sim3\sigma$) seen in $T_\mathrm{ASC}$ with $\Delta T_\mathrm{ASC}/P_\mathrm{b}\sim-7\times10^{-6}$. Hence, the orbital variations in PSR\,J0610$-$2100 are at least an order of magnitude smaller compared to those in PSRs\,B1957+20 and J2051$-$0827. 

In the spider timing model by \citet{Voisin19}, orbital variations are modelled by a gravitational quadrupole moment, which is split into spin $J_\mathrm{s}$, tidal $J_\mathrm{t}$, and time-dependent variable $J_v(t)$ components. The spin and tidal components are strong functions of the Roche-lobe radius, set by the mass ratio $q$ and Roche-lobe filling factor $f_\mathrm{RL}$. We compared $J_\mathrm{s}$ and $J_\mathrm{t}$ for PSRs\,J0610$-$2100 and J2051$-$0827. Assuming a 1.5\,M$_\odot$ pulsar mass and using the orbital inclinations of $i=73\degr$ and $i\sim40\degr$ for PSRs\,J0610$-$2100 and J2051$-$0827 \citep{Stappers01a} respectively, we used the mass function to obtain mass ratios. With $f_\mathrm{RL}=0.86$ for PSR\,J0610$-$2100 and $f_\mathrm{RL}=1$ for PSR\,J2051$-$0827 \citep{Stappers01a}, we found that these components are a factor 3.5 to 4 higher in PSR\,J2051$-$0827 compared to PSR\,J0610$-$2100. Unfortunately, we cannot compare the time-dependent variable components $J_v(t)$, as these depend on the observed orbital variations and are not linked to physical quantities in \citet{Voisin19}. It is unclear if the variable components scale similarly to the spin and tidal components of the quadrupole moment.

While the orbital period derivative obtained from timing itself is not a significant measurement, it is offset by the Shlovskii effect by $\dot{P}_\mathrm{b,shk}<7.91\times10^{-14}$\,s\,s$^{-1}$. At the best-fitting distance of 2.24\,kpc distance from light-curve modelling, we found that $\dot{P}_\mathrm{b,shk}=4.75\times10^{-14}$\,s\,s$^{-1}$, leading to an intrinsic orbital period derivative of $\dot{P}_\mathrm{b,int}=-12(3)\times10^{-14}$\,s\,s$^{-1}$. This tentative detection is an order of magnitude larger than the orbital period derivative expected due to emission of gravitational radiation, for which general relativity predicts $\dot{P}_\mathrm{b,GR}=-4.6\times10^{-15}$\,s\,s$^{-1}$ \citep{Peters64}. The presence of similar long-term orbital period derivatives in other black widow systems without showing detectable shorter term orbital period variations (e.g.\ PSR\,J0023+0923, \citealt{Bak20} and PSR\,J0636+5128, \citealt{Stovall14}) indicates that these may be the result of spin-orbit coupling as suggested by the models of \citet{Applegate94} and \citet{Voisin19}.

PSR\,J0610$-$2100 is currently timed as part of PTA experiments by the EPTA \citep{Desvignes16}, NANOgrav \cite{Arzoumanian18}, and the PPTA \citep{Kerr20}. Since the sensitivity of PTAs to gravitational waves is strongly dependent on the number of millisecond pulsars that can be timed to high precision \citep[e.g.][]{Siemens13}, continued timing of PSR\,J0610$-$2100 for PTA purposes is warranted. The absence of evidence for radio eclipses and significant orbital variations in the timing of PSR\,J0610$-$2100 confirms the suitability of this pulsar for PTAs. The long timing baseline of PSR\,J0610$-$2100 may provide clues whether the tentative detection of the orbital period derivative is a manifestation of spin-orbit coupling induced orbital variations that operate on time scales longer than those observed in systems such as PSRs\,B1957+20 and J2051$-$0827. This will help to assess the suitability of more recently discovered black widow systems for PTA purposes, even if optical observations show that the companion is irradiated by the millisecond pulsar.

\begin{acknowledgements}
    We thank the anonymous referee for the careful reading of and useful comments on the manuscript. 
    Based on observations collected at the European Organisation for Astronomical Research in the Southern Hemisphere under ESO programme 386.D-0207. This research was made possible by support from the Dutch National Science Agenda, NWA Startimpuls – 400.17.608.  C.~J.~C. and R.~P.~B acknowledge support from the ERC under the European Union's Horizon 2020 research and innovation programme (grant agreement No. 715051; Spiders). This work was supported by the Max-Planck-Gesellschaft~(MPG). The Nan\c{c}ay Radio Observatory is operated by the Paris Observatory, associated with the French Centre National de la Recherche Scientifique (CNRS). We acknowledge financial support from `Programme National de Cosmologie et Galaxies' (PNCG) of CNRS/INSU, France. The Westerbork Synthesis Radio Telescope is operated by the Netherlands Institute for Radio Astronomy (ASTRON) with support from The Netherlands Foundation for Scientific Research (NWO). The Parkes radio telescope is part of the Australia Telescope National Facility (grid.421683.a) which is funded by the Australian Government for operation as a National Facility managed by CSIRO. We acknowledge the Wiradjuri people as the traditional owners of the Observatory site. 
\end{acknowledgements}

\bibliographystyle{aa}
\bibliography{0610}

\end{document}